\documentclass[onecolumn]{svjour3}
\usepackage[sort&compress,numbers]{natbib}
\usepackage{subcaption,graphicx,float,caption, color}
\usepackage{amsmath,amsfonts,amssymb,mathrsfs,latexsym,esint}
\usepackage{yhmath}
\usepackage{stmaryrd}
\usepackage{placeins}
\usepackage{float}
\usepackage{mathtools}
\usepackage{tensor}
\usepackage{enumitem}
\usepackage{color}
\DeclareMathAlphabet{\mathsfit}{T1}{\sfdefault}{\mddefault}{\sldefault}
\SetMathAlphabet{\mathsfit}{bold}{T1}{\sfdefault}{\bfdefault}{\sldefault}

\def\bfu{{\bf u}}
\def\bfv{{\bf v}}
\def\bfx{{\bf x}}
\def\bfy{{\bf y}}

\def\bfB{{\bf B}}

\def\bfI{{\bf I}}

\def\bfL{{\bf L}}

\def\bfN{{\bf N}}

\def\bfS{{\bf S}}

\def\bfX{{\bf X}}

\def\bfe{{\bf e}}

\def\Atan{\mbox{\boldmath$\mathcal{A}$}}

\def\Ktan{\mbox{\boldmath$\mathcal{K}$}}
\def\Jtan{\mbox{\boldmath$\mathcal{J}$}}

\def\Atan{\mbox{\boldmath$\mathcal{A}$}}

\def\e0{\varepsilon_0}

\def\s0{\sigma_0}

\newcommand{\jump}[1]{\llbracket  #1 \rrbracket}

\long\def\symbolfootnote[#1]#2{\begingroup%
\def\thefootnote{\fnsymbol{footnote}}\footnote[#1]{#2}\endgroup}

\begin{document}

\journalname{}
\titlerunning{Homogenization of elastomers filled with liquid inclusions}

\title{Homogenization of elastomers filled with liquid inclusions: The small-deformation limit}

\author{Kamalendu Ghosh \and Victor Lef\`evre \and  Oscar Lopez-Pamies}

\institute{Kamalendu Ghosh \at Department of Civil and Environmental Engineering, University of Illinois, Urbana--Champaign, IL 61801-2352, USA \\
           \and
           Victor Lef\`evre \at Department of Mechanical Engineering, Northwestern University, Evanston, IL 60208, USA\\
           \and
          Oscar Lopez-Pamies \at
           Department of Civil and Environmental Engineering, University of Illinois, Urbana--Champaign, IL 61801-2352, USA \\
           D\'epartement de M\'ecanique, \'Ecole Polytechnique, 91128 Palaiseau, France\\
           \email{pamies@illinois.edu}
           }

\date{Received: date / Accepted: date}
\maketitle

\begin{abstract}

This paper presents the derivation of the homogenized equations that describe the macroscopic mechanical response of elastomers filled with liquid inclusions in the setting of small quasistatic deformations. The derivation is carried out for materials with periodic microstructure by means of a two-scale asymptotic analysis. The focus is on the non-dissipative case when the elastomer is an elastic solid, the liquid making up the inclusions is an elastic fluid, the interfaces separating the solid elastomer from the liquid inclusions are elastic interfaces featuring an initial surface tension, and the inclusions are initially $n$-spherical ($n=2,3$) in shape. Remarkably, in spite of the presence of local residual stresses within the inclusions due to an initial surface tension at the interfaces, the macroscopic response of such filled elastomers turns out to be that of a linear elastic solid that is free of residual stresses and hence one that is simply characterized by an effective modulus of elasticity $\overline{\bfL}$. What is more, in spite of the fact that the local moduli of elasticity in the bulk and the interfaces do not possess minor symmetries (due to the presence of residual stresses and the initial surface tension at the interfaces), the resulting effective modulus of elasticity $\overline{\bfL}$ does possess the standard minor symmetries of a conventional linear elastic solid, that is, $\overline{L}_{ijkl}=\overline{L}_{jikl}=\overline{L}_{ijlk}$. As an illustrative application, numerical results are worked out and analyzed for the effective modulus of elasticity of isotropic suspensions of incompressible liquid $2$-spherical inclusions of monodisperse size embedded in an isotropic incompressible elastomer.

\keywords{Suspensions; Size effects; Metamaterials; Multiscale asymptotic expansions}

\end{abstract}

\section{Introduction}\label{Sec:Intro}

{\color{black} A series of experimental and theoretical investigations of late have pointed to elastomers filled with \emph{liquid} inclusions --- contrary to conventional \emph{solid} fillers --- as a new class of materials with unique macroscopic mechanical/physical properties \cite{LP14,Syleetal15,Bartlettetal2017,LDLP17,LGLP19,Yunetal19}. Two reasons are behind such properties.

The first is that the addition of liquid inclusions to elastomers increases the overall deformability. This is in contrast to the addition of conventional fillers, which, being typically made of stiff solids, decreases deformability. Magnetorheological elastomers (MREs) are a class of materials that makes this dichotomy readily apparent. For instance, while MREs filled with iron particles are able to undergo very modest deformations even when subjected to large magnetic fields, MREs filled with ferrofluid inclusions are able to undergo significant deformations when subjected to modest magnetic fields. This is because of the increased deformability imparted by the ferrofluid inclusions compared to that of iron particles \cite{LDLP17,LGLP19}.

The second reason behind the fascinating properties of elastomers filled with liquid inclusions is that the behavior of the interfaces separating a solid elastomer from embedded liquid inclusions feature their own mechanical/physical behavior, one that, while negligible when the inclusions are ``large'', may dominate the macroscopic properties of the material when the inclusions are sufficiently ``small''. The experiments on a silicone elastomer filled with ionic-liquid droplets reported in \cite{Syleetal15} provide a recent visual example of this size-dependent phenomenon. Precisely, these experiments show that, under the same applied mechanical loads, droplets with smaller radii undergo significantly smaller deformations. This is because smaller droplets feature a larger interface stiffness --- or, more specifically, a larger initial elasto-capillary number $eCa$ --- that scales inversely proportional with their radius (see Section \ref{Sec: Application} below).

While the above twofold \emph{qualitative} understanding is well settled, a \emph{quantitative} understanding of the mechanics of elastomers filled with liquid inclusions is yet to be fully developed. In this context, Ghosh and Lopez-Pamies \cite{GLP22} have recently worked out several theoretical results aimed at explaining and describing the mechanics of deformation of elastomers embedding liquid inclusions. \emph{Inter alia}, these include the homogenized equations that, in the basic setting of small quasistatic elastic deformations, describe the macroscopic mechanical response of elastomers filled with liquid inclusions that are initially spherical in shape (see Section 3 in \cite{GLP22}). The objective of this paper is to present the derivation of this homogenization limit. The derivation focuses on materials with periodic microstructure and is carried out by means of a two-scale asymptotic analysis.
}

\section{The problem}\label{Sec:The problem}

Consider the boundary-value problem
\begin{align}
\left\{\begin{array}{l}
 {\rm Div}\left[\bfL^\epsilon(\bfX)\nabla\bfu^\epsilon\right]=\textbf{0},\quad \bfX\in {\rm \Omega}\setminus{\rm \Gamma}\vspace{0.2cm}\\
  \widehat{{\rm Div}}\left[\widehat{\bfL}^\epsilon\widehat{\nabla}\bfu^\epsilon\right]-\jump{\bfL^\epsilon(\bfX)\nabla\bfu^\epsilon}\widehat{\bfN}=\textbf{0},\quad \bfX\in {\rm \Gamma}\vspace{0.2cm}\\
\bfu^{\epsilon}(\bfX)=\overline{\bfu}(\bfX),\quad \bfX\in\partial{\rm \Omega}\end{array}\right.  \label{BVP_epsilon}
\end{align}
for the displacement field $\bfu^\epsilon(\bfX)\in H^1 ({\rm \Omega};\mathbb{R}^n)$ in an open domain $\mathrm{\Omega}\subset \mathbb{R}^n$, $n=2,3$, with boundary $\partial\mathrm{\Omega}$. Ghosh and Lopez-Pamies \cite{GLP22} have recently shown that (\ref{BVP_epsilon}) are the equations that govern the mechanical response of an elastomeric matrix ($\texttt{m}$) filled with initially $n$-spherical\footnote{Employing the parlance of geometers (\cite{Coxeter73}, Section 7.3), we refer to circles as $2$-spheres and to spheres as $3$-spheres.} liquid inclusions ($\texttt{i}$) of length scale $\epsilon$ subjected to small quasistatic deformations. Here, $\bfL^\epsilon(\bfX)$ stands for the modulus of elasticity for the bulk ${\rm \Omega}\setminus{\rm \Gamma}$, which is comprised of the solid elastomeric matrix and the firmly embedded liquid inclusions, $\widehat{\bfL}^\epsilon$  denotes the modulus of elasticity for the interfaces ${\rm \Gamma}$ separating the elastomer from the inclusions, $\widehat{\bfN}$ is the unit normal of ${\rm \Gamma}$ pointing outwards from the inclusions towards the elastomer, and $\overline{\bfu}(\bfX)$ is the applied displacement boundary condition (Dirichlet boundary conditions are assumed for simplicity of presentation). In equations (\ref{BVP_epsilon}), ${\rm Div}$ stands for the bulk divergence operator, $\jump{\cdot}$ is the jump operator across the interfaces ${\rm \Gamma}$ based on the convention $\jump{f(\bfX)}=f^{(\texttt{i})}(\bfX)-f^{(\texttt{m})}(\bfX)$, where $f^{(\texttt{i})}$ (resp. $f^{(\texttt{m})}$) denotes the limit of any given function $f(\bfX)$ when approaching ${\rm \Gamma}$ from within the inclusion (resp. matrix), while $\widehat{\nabla}$ and $\widehat{{\rm Div}}$ stand for the interface gradient and divergence operators. In indicial notation, with respect to a Cartesian frame of reference $\{\bfe_i\}$ ($i=1,...,n$) and help of the projection tensor
\begin{equation*}
\widehat{\bfI}=\bfI - \widehat{\bfN} \otimes \widehat{\bfN},
\end{equation*}
we recall that these interface operators read \cite{Carmo16,GWL98}
\begin{equation*}
\left(\widehat{\nabla}\bfv\right)_{ij}=\dfrac{\partial v_i}{\partial X_k}(\bfX)\widehat{I}_{kj}\qquad {\rm and}\qquad
\left(\widehat{{\rm Div}}\,\bfS\right)_{i}=\dfrac{\partial S_{ij}}{\partial X_k}(\bfX)\widehat{I}_{kj},\qquad \bfX \in{\rm \Gamma}
\end{equation*}
when applied to vector and second-order tensor fields.

\begin{figure}
\begin{center}
\includegraphics[width=4.4in]{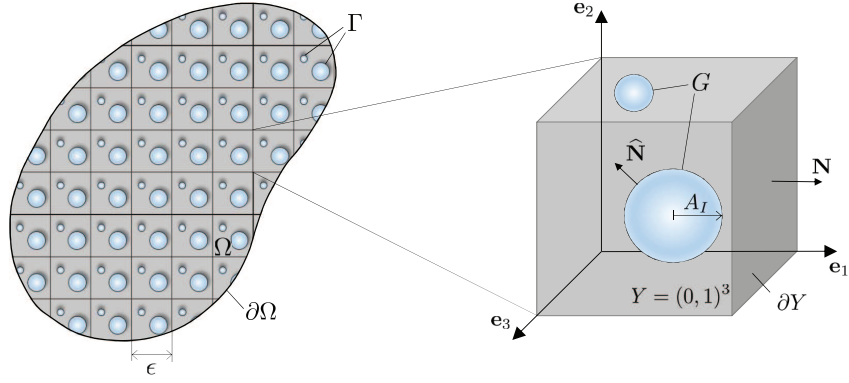}
   \vspace{-0.2cm}
   \caption{\small Schematics of the initial configuration ${\rm \Omega}\in \mathbb{R}^3$ of a periodic suspension, of period $\epsilon$, of $3$-spherical liquid inclusions embedded in a solid elastomer and of its defining unit cell $Y=(0,1)^3$. The radii of the inclusions are denoted by $A_I^\epsilon=\epsilon A_I$ and their outward unit normal by $\widehat{\bfN}$. The interfaces separating the elastomer from the inclusions in ${\rm \Omega}$ are denoted by ${\rm \Gamma}$. Within the unit cell $Y$, the interfaces separating the elastomer from the inclusions are denoted by $G$.}
 \label{Fig1}
 \end{center}
 \vspace{-0.2cm}
\end{figure}

\paragraph{Filled elastomers with periodic microstructure} For filled elastomers with periodic microstructure, which are the class of materials of interest in this work, the initial subdomains occupied collectively by all the inclusions can be expediently described by the characteristic function
\begin{equation}
\theta^\epsilon(\bfX)=\displaystyle\sum_{I=1}^{\texttt{N}}\theta^\epsilon_I(\bfX)\label{theta}
\end{equation}
in terms of the characteristic functions
\begin{equation}
\theta^\epsilon_I(\bfX)=\theta_I(\epsilon^{-1}\bfX)\qquad I=1,...,\texttt{N} \label{thetaj}
\end{equation}
for each individual inclusion. Here, $\theta_I(\bfy)$ are $ Y$-periodic functions, with $ Y=(0,1)^n$, and $\texttt{N}$ denotes the number of inclusions contained in the unit cell $Y$. It immediately follows that $\theta^\epsilon(\bfX)=\theta(\epsilon^{-1}\bfX)$, where $\theta(\bfy)$ is also $ Y$-periodic. Figure \ref{Fig1} shows a schematic of a filled elastomer with periodic microstructure in its initial configuration for an illustrative case of space dimension $n=3$ and $\texttt{N}=2$ inclusions in $Y$.
%
%

Granted (\ref{theta})-(\ref{thetaj}), the modulus of elasticity for the bulk and the interfaces read, respectively, as
\begin{align}
\bfL^\epsilon(\bfX)=\left(1-\theta(\epsilon^{-1}\bfX)\right)\bfL^{(\texttt{m})}+
\displaystyle\sum_{I=1}^{\texttt{N}}\theta_I(\epsilon^{-1}\bfX)\left[ n \Lambda^{(\texttt{i})}\Jtan+r^\epsilon_{I}(\Atan-\Ktan+(n-1)\Jtan)\right]\label{Lbulk}
\end{align}
and
\begin{equation}
\widehat{\bfL}^\epsilon=2\,\widehat{\mu}^{\,\epsilon}\,\widehat{\Ktan}+2 (\widehat{\mu}^{\,\epsilon}+\widehat{\Lambda}^{\,\epsilon})\widehat{\Jtan}+
\widehat{\gamma}^{\epsilon}\left(\widehat{\Atan}-\widehat{\Ktan}+\widehat{\Jtan}\right).\label{Linterface}
\end{equation}
In relation (\ref{Lbulk}), $\Atan$, $\Ktan$, $\Jtan$ are the orthonormal\footnote{That is, $\Atan\Ktan=\Ktan\Atan=\Atan\Jtan=\Jtan\Atan=\Ktan\Jtan=\Jtan\Ktan=\textbf{0}$, $\Atan\Atan=\Atan$, $\Ktan\Ktan=\Ktan$, and $\Jtan\Jtan=\Jtan$.} eigentensors
\begin{align}
&\mathcal{A}_{ijkl}=\dfrac{1}{2}(\delta_{ik}\delta_{jl}-\delta_{il}\delta_{jk}),\label{A}\\
& \mathcal{K}_{ijkl}=\dfrac{1}{2}\left(\delta_{ik}\delta_{jl}+\delta_{il}\delta_{jk}\right)-\dfrac{1}{n}\delta_{ij}\delta_{kl},\label{K}\\
&\mathcal{J}_{ijkl}=\dfrac{1}{n}\delta_{ij}\delta_{kl},\label{J}
\end{align}
$\bfL^{(\texttt{m})}$ is the modulus of elasticity of the elastomeric matrix, which satisfies the standard symmetry and positive-definiteness properties
\begin{align*}
L^{(\texttt{m})}_{ijkl}=L^{(\texttt{m})}_{klij}=L^{(\texttt{m})}_{jikl}=L^{(\texttt{m})}_{ijlk},\qquad  B_{ij}L^{(\texttt{m})}_{ijkl}B_{kl}\geq \alpha  B_{pq}  B_{pq}\;\forall \bfB\in \mathbb{R}^{n\times n}
\end{align*}
and some $\alpha>0$, $\Lambda^{(\texttt{i})}\geq 0$ is the first Lam\'e constant of the liquid making up the inclusions, and
\begin{align}
r^\epsilon_{I}=-\dfrac{(n-1)\,\widehat{\gamma}^\epsilon}{A^\epsilon_I}\qquad I=1,...,\texttt{N}\label{rj}
\end{align}
denotes the initial hydrostatic stress that the $I$th inclusion is subjected to in the initial configuration. In this last expression, $\widehat{\gamma}^\epsilon$ stands for the initial surface tension on the interfaces and
\begin{align*}
A_I^\epsilon=\epsilon A_I
\end{align*}
is the radius of the $I$th inclusion, where $0<A_I<1$. In relation (\ref{Linterface}), $\widehat{\Atan}$, $\widehat{\Ktan}$, $\widehat{\Jtan}$ are the orthonormal\footnote{In complete analogy with their bulk counterparts (\ref{A})-(\ref{J}), $\widehat{\Atan}\widehat{\Ktan}=\widehat{\Ktan}\widehat{\Atan}=\widehat{\Atan}\widehat{\Jtan}=\widehat{\Jtan}\widehat{\Atan}=\widehat{\Ktan}\widehat{\Jtan}=\widehat{\Jtan}\widehat{\Ktan}=\textbf{0}$,
$\widehat{\Atan}\widehat{\Atan}=\widehat{\Atan}$, $\widehat{\Ktan}\widehat{\Ktan}=\widehat{\Ktan}$, and $\widehat{\Jtan}\widehat{\Jtan}=\widehat{\Jtan}$.} eigentensors
\begin{align*}
&\widehat{\mathcal{A}}_{ijkl}=\delta_{ik}\widehat{I}_{jl}-\dfrac{1}{2}\left(\widehat{I}_{ik}\widehat{I}_{jl}+\widehat{I}_{il}\widehat{I}_{jk}\right),\nonumber\\
&\widehat{\mathcal{K}}_{ijkl}=\dfrac{1}{2}\left(\widehat{I}_{ik}\widehat{I}_{jl}+\widehat{I}_{il}\widehat{I}_{jk}-\widehat{I}_{ij}\widehat{I}_{kl}\right),\nonumber\\
&\widehat{\mathcal{J}}_{ijkl}=\dfrac{1}{2}\widehat{I}_{ij}\widehat{I}_{kl},\label{AKJ-hat}
\end{align*}
$\widehat{\mu}^{\,\epsilon}\geq 0$ and $\widehat{\Lambda}^{\,\epsilon}\geq 0$ are the interface Lam\'{e} constants, and, again, $\widehat{\gamma}^\epsilon\geq 0$ denotes the surface tension on the interfaces in the initial configuration.

\begin{remark}
{\rm All inclusions are assumed to be made of the same liquid, thus the unique value of Lam\'{e} constant $\Lambda^{(\texttt{i})}$ in (\ref{Lbulk}); note that  the case of an incompressible liquid is recovered by setting $\Lambda^{(\texttt{i})}=+\infty$. However, because each inclusion is allowed to have its own initial size, the residual hydrostatic stresses $r^\epsilon_{I}$ in (\ref{Lbulk}) may be different for different inclusions.}
\end{remark}

\begin{remark}
{\rm The specific form of the residual hydrostatic stresses (\ref{rj}) is necessarily a direct consequence of equilibrium within the bulk of the liquid making up the inclusions and on the interfaces separating the inclusions from the elastomer in the initial configuration. Indeed, the residual hydrostatic stresses (\ref{rj}) are the solutions of the equations
\begin{equation}
\left\{\begin{array}{l}\nabla\left[ \displaystyle\sum_{I=1}^{\texttt{N}}\theta_j(\epsilon^{-1}\bfX) r^\epsilon_I\right]={\bf0}, \quad  \bfX\in\mathrm{\Omega}\setminus {\rm \Gamma} \vspace{0.4cm}\\
\displaystyle\sum_{I=1}^{\texttt{N}}\theta_I(\epsilon^{-1}\bfX) r^\epsilon_I=-\widehat{\gamma}^\epsilon\;{\rm tr}\,\widehat{\nabla}\widehat{\bfN}, \quad  \bfX\in{\rm \Gamma}
\end{array}\right. .\label{BLM-Young-Laplace-0}
\end{equation}
Remark that the first of these equations is nothing more than balance of linear momentum within the inclusions, while the second one is the Young-Laplace equation.
  }
\end{remark}

\paragraph{Scaling of the interface Lam\'{e} constants $\widehat{\mu}^{\,\epsilon}$, $\widehat{\Lambda}^{\,\epsilon}$ and initial surface tension $\widehat{\gamma}^\epsilon$} The governing equations (\ref{BVP_epsilon}), with (\ref{Lbulk}), (\ref{Linterface}), and (\ref{rj}), apply to elastomers filled with a periodic distribution of spherical liquid inclusions of arbitrary length scale $\epsilon$. In this work, we are interested in the limit as $\epsilon\searrow 0$ when the inclusions are much smaller that the length scale of ${\rm \Omega}$, which is considered to be a fixed domain. To this end, remark that equations (\ref{BVP_epsilon}) depend directly on the size of the inclusions through the residual hydrostatic stresses (\ref{rj}) in (\ref{BVP_epsilon})$_{1,2}$ and through the interface divergence $\widehat{{\rm Div}}$ operator in (\ref{BVP_epsilon})$_{2}$. Accordingly, in order to preserve the correct physics in the limit as $\epsilon\searrow 0$, the interface Lam\'{e} constants $\widehat{\mu}^{\,\epsilon}$, $\widehat{\Lambda}^{\,\epsilon}$ and initial surface tension $\widehat{\gamma}^\epsilon$ must scale appropriately with $\epsilon$, in particular, they must scale linearly in $\epsilon$. We write
\begin{align}
\widehat{\mu}^{\epsilon}=\epsilon\, \widehat{\mu},\quad \widehat{\Lambda}^{\epsilon}=\epsilon\, \widehat{\Lambda},\quad \widehat{\gamma}^\epsilon=\epsilon\, \widehat{\gamma},\label{gamma-eps}
\end{align}
where $\widehat{\mu}\geq0$, $\widehat{\Lambda}\geq0$, and $\widehat{\gamma}\geq0$.

Granted the scaling (\ref{gamma-eps}), the modulus of elasticity (\ref{Lbulk}) for the bulk  depends on $\epsilon$ only through the combination $\epsilon^{-1}\bfX$, specifically,
\begin{align}
\bfL^\epsilon(\bfX)=&\left(1-\theta(\epsilon^{-1}\bfX)\right)\bfL^{(\texttt{m})}+\nonumber\\
&\displaystyle\sum_{I=1}^{\texttt{N}}\theta_I(\epsilon^{-1}\bfX)\left[n\Lambda^{(\texttt{i})}\Jtan-\dfrac{(n-1)\widehat{\gamma}}{A_I}(\Atan-\Ktan+(n-1)\Jtan)\right]=: \,\bfL(\epsilon^{-1}\bfX),\label{Lbulk-eps}
\end{align}
while the modulus of elasticity (\ref{Linterface}) for the interfaces specializes to
\begin{align}
\widehat{\bfL}^\epsilon= \epsilon \left( 2\,\widehat{\mu}\,\widehat{\Ktan}+2 (\widehat{\mu}+\widehat{\Lambda})\widehat{\Jtan}+
\widehat{\gamma} \left(\widehat{\Atan}-\widehat{\Ktan}+\widehat{\Jtan}\right)\right)=:\epsilon\,\widehat{\bfL}.\label{Linterface-eps}
\end{align}
It follows that the boundary-value problem (\ref{BVP_epsilon}) specializes to
\begin{align}
\left\{\begin{array}{l}
 {\rm Div}\left[\bfL(\epsilon^{-1}\bfX)\nabla\bfu^\epsilon\right]=\textbf{0},\quad \bfX\in {\rm \Omega}\setminus{\rm \Gamma}\vspace{0.2cm}\\
  \widehat{{\rm Div}}\left[\epsilon\,\widehat{\bfL}\widehat{\nabla}\bfu^\epsilon\right]-\jump{\bfL(\epsilon^{-1}\bfX)\nabla\bfu^\epsilon}\widehat{\bfN}=\textbf{0},\quad \bfX\in {\rm \Gamma}\vspace{0.2cm}\\
\bfu^{\epsilon}(\bfX)=\overline{\bfu}(\bfX),\quad \bfX\in\partial{\rm \Omega}\end{array}\right. .  \label{BVP-1}
\end{align}

For fixed $\epsilon$, equations (\ref{BVP-1}) generalize in two counts the classical linear elastostatics equations for heterogeneous materials. Specifically, these equations feature: ($i$) residual stresses (in the inclusions) and ($ii$) a non-standard jump condition across material (matrix/inclusions) interfaces due to the presence of interfacial forces. These two traits have profound implications not only on the resulting mechanical response of the body, but also on the mathematical analysis of the problem. Indeed, remark that the \emph{non-symmetric} term
\begin{equation*}
-\theta_I(\epsilon^{-1}\bfX)\dfrac{(n-1)\widehat{\gamma}}{A_I}\Atan
\end{equation*}
in (\ref{Lbulk-eps}) makes the bulk modulus of elasticity $\bfL(\epsilon^{-1}\bfX)$ \emph{not} positive definite. Similarly, for the physically prominent case when $\widehat{\gamma}>\widehat{\mu}$, the \emph{negative} term
\begin{equation*}
-\widehat{\gamma}\,\widehat{\Ktan}
\end{equation*}
in (\ref{Linterface-eps}) makes the interface modulus of elasticity $\widehat{\bfL}$ \emph{not} positive definite. Accordingly, the standard coercivity based on local positive definiteness cannot be invoked here to prove existence of solution for (\ref{BVP-1}) via the Lax-Milgram theorem. Nevertheless, the expectation\footnote{In point of fact, explicit solutions can be readily worked out in terms of plane/spherical harmonics for some special cases, see, e.g., \cite{Sharmaetal03,Syleetal15b,GLP22}.} is that one can identify an appropriate weaker notion of coercivity that allows to prove existence. We shall address this issue in a separate contribution. From now onward, we simply assume that solutions $\bfu^\epsilon(\bfX)\in H^1 ({\rm \Omega};\mathbb{R}^n)$ exist for (\ref{BVP-1}).

\section{The limit as $\epsilon\searrow 0$ by the method of two-scale asymptotic expansions}\label{Sec: Expansion}

In this section, we present the derivation of the homogenized equations that emerge from the boundary-value problem (\ref{BVP-1}) in the limit as $\epsilon\searrow 0$ by means of the method of two-scale asymptotic expansions \cite{SP80,BLP11}.

We begin by looking for solutions of the asymptotic form
\begin{align}
u_i^\epsilon(\bfX)=& u_i^{(0)}(\bfX,\epsilon^{-1}\bfX)+\epsilon\,u_i^{(1)}(\bfX,\epsilon^{-1}\bfX)+\epsilon^2 u_i^{(2)}(\bfX,\epsilon^{-1}\bfX)+...\nonumber\\
=&\sum_{s=0}^{\infty}\epsilon^s u_i^{(s)}(\bfX,\epsilon^{-1}\bfX), \label{ansatz}
\end{align}
where the functions $\bfu^{(s)}(\bfX,\epsilon^{-1}\bfX)$ are $ Y$-periodic in their second argument and, according to the boundary condition (\ref{BVP-1})$_3$, such that $\bfu^{(0)}(\bfX,\epsilon^{-1}\bfX)=\overline{\bfu}(\bfX)$ and $\bfu^{(s)}(\bfX,\epsilon^{-1}\bfX)=\bf0$ for $s\neq 0$ on $\partial {\rm \Omega}$.

Next, we introduce the variables
\begin{equation*}
\bfx=\bfX\quad {\rm and}\quad \bfy=\epsilon^{-1}\bfX
\end{equation*}
and operators
\begin{align*}
\texttt{A}_{ik}^\epsilon=\epsilon^{-2}\texttt{A}_{ik}^{(1)}+\epsilon^{-1}\texttt{A}_{ik}^{(2)}+\texttt{A}_{ik}^{(3)}&\quad {\rm with}\nonumber\\
&\texttt{A}_{ik}^{(1)}=\dfrac{\partial}{\partial y_j}\left[L_{ijkl}\left(\bfy\right)\dfrac{\partial }{\partial y_l}\right],\nonumber\\
&\texttt{A}_{ik}^{(2)}=\dfrac{\partial}{\partial y_j}\left[L_{ijkl}\left(\bfy\right)\dfrac{\partial }{\partial x_l}\right]+\dfrac{\partial}{\partial x_j}\left[L_{ijkl}\left(\bfy\right)\dfrac{\partial }{\partial y_l}\right],\nonumber\\
&\texttt{A}_{ik}^{(3)}=\dfrac{\partial}{\partial x_j}\left[L_{ijkl}\left(\bfy\right)\dfrac{\partial }{\partial x_l}\right],
\end{align*}
and
\begin{align*}
\widehat{\texttt{A}}_{ik}^{\,\epsilon}=&\epsilon^{-1}\widehat{\texttt{A}}_{ik}^{(1)}+\widehat{\texttt{A}}_{ik}^{(2)}+\epsilon\,\widehat{\texttt{A}}_{ik}^{(3)}\quad {\rm with}\nonumber\\
&\widehat{\texttt{A}}_{ik}^{(1)}=\dfrac{\partial}{\partial y_q}\left[\widehat{L}_{ijkl}\dfrac{\partial }{\partial y_p} \widehat{I}_{pl}\right]\widehat{I}_{qj}-\jump{L_{ijkl}(\bfy)\dfrac{\partial }{\partial y_l}}\widehat{N}_j,\nonumber\\
&\widehat{\texttt{A}}_{ik}^{(2)}=\dfrac{\partial}{\partial y_q}\left[\widehat{L}_{ijkl} \dfrac{\partial }{\partial x_p} \widehat{I}_{pl}\right]\widehat{I}_{qj}+\dfrac{\partial}{\partial x_q}\left[\widehat{L}_{ijkl} \dfrac{\partial }{\partial y_p} \widehat{I}_{pl}\right] \widehat{I}_{qj}-\jump{L_{ijkl}(\bfy)\dfrac{\partial }{\partial x_l} }\widehat{N}_j,\nonumber\\
&\widehat{\texttt{A}}_{ik}^{(3)}=\dfrac{\partial}{\partial x_q}\left[\widehat{L}_{ijkl} \dfrac{\partial }{\partial x_p} \widehat{I}_{pl}\right]\widehat{I}_{qj},
\end{align*}
in terms of which equations (\ref{BVP-1})$_{1,2}$ can be compactly rewritten as
\begin{align}
\left\{\begin{array}{l}
 \texttt{A}_{ik}^{\epsilon} u_k^\epsilon=0\vspace{0.2cm}\\
 \widehat{\texttt{A}}_{ik}^{\,\epsilon} u_k^\epsilon=0\end{array}\right. .  \label{PDEs-1}
\end{align}

Substituting the ansatz (\ref{ansatz}) in the PDEs (\ref{PDEs-1}) and expanding in powers of $\epsilon$ leads to a hierarchy of equations for the functions ${\bfu}^{(s)}(\bfx,\bfy)$. Only the first four of these, of $O(\epsilon^{-2})$, $O(\epsilon^{-1})$, $O(\epsilon^{0})$, and $O(\epsilon)$, turn out to be needed for our purposes here. In terms of the above-introduced operators, they read
\begin{align}
&\texttt{A}_{ik}^{(1)}u^{(0)}_k=0, \label{Asymptotic Eq1}\\
&\left\{\begin{array}{l}
\texttt{A}_{ik}^{(1)}u^{(1)}_k+\texttt{A}_{ik}^{(2)}u^{(0)}_k=0\vspace{0.2cm}\\
\widehat{\texttt{A}}_{ik}^{(1)}u^{(0)}_k=0
\end{array}\right. ,\label{Asymptotic Eq2}\\
&\left\{\begin{array}{l}
\texttt{A}_{ik}^{(1)}u^{(2)}_k+\texttt{A}_{ik}^{(2)}u^{(1)}_k+\texttt{A}_{ik}^{(3)}u^{(0)}_k=0\vspace{0.2cm}\\
\widehat{\texttt{A}}_{ik}^{(1)}u^{(1)}_k+\widehat{\texttt{A}}_{ik}^{(2)}u^{(0)}_k=0
\end{array}\right. ,\label{Asymptotic Eq3}\\
&\left\{\begin{array}{l}
\texttt{A}_{ik}^{(1)}u^{(3)}_k+\texttt{A}_{ik}^{(2)}u^{(2)}_k+\texttt{A}_{ik}^{(3)}u^{(1)}_k=0\vspace{0.2cm}\\
\widehat{\texttt{A}}_{ik}^{(1)}u^{(2)}_k+\widehat{\texttt{A}}_{ik}^{(2)}u^{(1)}_k+\widehat{\texttt{A}}_{ik}^{(3)}u^{(0)}_k=0
\end{array}\right. . \label{Asymptotic Eq4}
\end{align}

\paragraph{\textbf{The equations of $O(\epsilon^{-2})$ in the bulk and $O(\epsilon^{-1})$ on the interfaces}} The PDEs (\ref{Asymptotic Eq1}) and (\ref{Asymptotic Eq2})$_2$ can be combined to render the set of equations
\begin{align}
\left\{\begin{array}{l}
\dfrac{\partial}{\partial y_j}\left[L_{ijkl}\left(\bfy\right)\dfrac{\partial u^{(0)}_k}{\partial y_l}(\bfx,\bfy)\right]=0,\quad\bfy\in Y\setminus G\vspace{0.4cm}\\
\dfrac{\partial}{\partial y_q}\left[\widehat{L}_{ijkl}\dfrac{\partial u^{(0)}_k}{\partial y_p}(\bfx,\bfy) \widehat{I}_{pl}\right]\widehat{I}_{qj}-\jump{L_{ijkl}(\bfy)\dfrac{\partial u^{(0)}_k}{\partial y_l}(\bfx,\bfy)}\widehat{N}_j=0,\quad\bfy\in G
\end{array}\right.\label{Eq-u0}
\end{align}
for the function $\bfu^{(0)}_k(\bfx,\bfy)$ in the unit cell $Y$, where $G$ has been introduced to denote the interfaces separating the elastomer from the inclusions contained in $Y$. In (\ref{Eq-u0}), $\bfy$ plays the role of the independent variable, whereas $\bfx$ is just a parameter. Accordingly, the solution of (\ref{Eq-u0}) with respect to $\bfy$ is simply a function of $\bfx$ that does not depend on $\bfy$. We write
\begin{equation}
\bfu^{(0)}(\bfx,\bfy)=\bfu(\bfx).\label{Sol-u0}
\end{equation}

\paragraph{\textbf{The equations of $O(\epsilon^{-1})$ in the bulk and $O(\epsilon^{0})$ on the interfaces}} Making direct use of the result (\ref{Sol-u0}), the PDEs
(\ref{Asymptotic Eq2})$_1$ and (\ref{Asymptotic Eq3})$_2$ can be combined to yield
\begin{align}
\left\{\begin{array}{l}
\dfrac{\partial}{\partial y_j}\left[L_{ijkl}\left(\bfy\right)\dfrac{\partial u^{(1)}_k}{\partial y_l}(\bfx,\bfy)\right]=-\dfrac{\partial}{\partial y_j}\left[L_{ijkl}\left(\bfy\right)\dfrac{\partial u_k}{\partial x_l}(\bfx)\right],\quad\bfy\in Y\setminus G\vspace{0.4cm}\\
\dfrac{\partial}{\partial y_q}\left[\widehat{L}_{ijkl}\dfrac{\partial u^{(1)}_k}{\partial y_p}(\bfx,\bfy) \widehat{I}_{pl}\right]\widehat{I}_{qj}-\jump{L_{ijkl}(\bfy)\dfrac{\partial u^{(1)}_k}{\partial y_l}(\bfx,\bfy)}\widehat{N}_j=\vspace{0.2cm}\\
-\dfrac{\partial}{\partial y_q}\left[\widehat{L}_{ijkl} \dfrac{\partial u_k}{\partial x_p}(\bfx) \widehat{I}_{pl}\right]\widehat{I}_{qj}+\jump{L_{ijkl}(\bfy)\dfrac{\partial u_k}{\partial x_l}(\bfx)}\widehat{N}_j,\quad\bfy\in G
\end{array}\right. ,\label{Eq-u1}
\end{align}
which, for a given function $\bfu(\bfx)$, can be thought of as equations for the function $\bfu^{(1)}(\bfx,\bfy)$ in the unit cell $Y$ with $\bfx$ playing the role of a parameter.

By introducing the $Y$-periodic function $\omega_{kmn}(\bfy)\in H^1(Y;\mathbb{R}^{n^{3}})$ defined implicitly as the solution of the unit-cell problem
\begin{align}
\left\{\begin{array}{l}
\dfrac{\partial}{\partial y_j}\left[L_{ijkl}\left(\bfy\right)\dfrac{\partial \omega_{kmn}}{\partial y_l}(\bfy)\right]=-\dfrac{\partial L_{ijmn}}{\partial y_j}\left(\bfy\right),\quad\bfy\in Y\setminus G\vspace{0.4cm}\\
\dfrac{\partial}{\partial y_q}\left[\widehat{L}_{ijkl}\dfrac{\partial \omega_{kmn}}{\partial y_p}(\bfy) \widehat{I}_{pl}\right]\widehat{I}_{qj}-\jump{L_{ijkl}(\bfy)\dfrac{\partial \omega_{kmn}}{\partial y_l}(\bfy)}\widehat{N}_j=\vspace{0.2cm}\\
-\dfrac{\partial}{\partial y_q}\left[\widehat{L}_{ijkl}\delta_{km} \widehat{I}_{nl}\right]\widehat{I}_{qj}+\jump{L_{ijkl}(\bfy)\delta_{km}\delta_{ln}}\widehat{N}_j,\quad\bfy\in G\vspace{0.4cm}\\
\int_{Y}\omega_{kmn}(\bfy){\rm d}\bfy=0
\end{array}\right. ,\label{Eq-omega}
\end{align}
the solution (with respect to $\bfy$) of (\ref{Eq-u1}) can be written in the separable form
\begin{align}
u_k^{(1)}(\bfx,\bfy)=\omega_{kmn}(\bfy)\dfrac{\partial u_{m}}{\partial x_n}(\bfx)+v^{(1)}_k(\bfx),\label{Sol-u1}
\end{align}
where $\bfv^{(1)}(\bfx)$ is an arbitrary function of $\bfx$.

\paragraph{\textbf{The equations of $O(\epsilon^{0})$ in the bulk and $O(\epsilon)$ on the interfaces}} In turn, making again use of the result (\ref{Sol-u0}), the combination of PDEs (\ref{Asymptotic Eq3})$_1$ and (\ref{Asymptotic Eq4})$_2$ renders the set of equations
\begin{align}
\left\{\begin{array}{l}
\dfrac{\partial}{\partial y_j}\left[L_{ijkl}\left(\bfy\right)\dfrac{\partial u^{(2)}_k}{\partial y_l}(\bfx,\bfy)\right]=-\dfrac{\partial}{\partial y_j}\left[L_{ijkl}\left(\bfy\right)\dfrac{\partial u^{(1)}_k}{\partial x_l}(\bfx,\bfy)\right]-\vspace{0.2cm}\\
\hspace{3.2cm}\dfrac{\partial}{\partial x_j}\left[L_{ijkl}(\bfy)\left(\dfrac{\partial u_k}{\partial x_l}(\bfx)+\dfrac{\partial u^{(1)}_k}{\partial y_l}(\bfx,\bfy)\right)\right],\quad\bfy\in Y\setminus G\vspace{0.4cm}\\
\dfrac{\partial}{\partial y_q}\left[\widehat{L}_{ijkl}\dfrac{\partial u^{(2)}_k}{\partial y_p}(\bfx,\bfy) \widehat{I}_{pl}\right]\widehat{I}_{qj}-\jump{L_{ijkl}(\bfy)\dfrac{\partial u^{(2)}_k}{\partial y_l}(\bfx,\bfy)}\widehat{N}_j=\vspace{0.2cm}\\
\hspace{0.45cm}-\dfrac{\partial}{\partial y_q}\left[\widehat{L}_{ijkl} \dfrac{\partial u^{(1)}_k}{\partial x_p}(\bfx,\bfy) \widehat{I}_{pl}\right]\widehat{I}_{qj}+\jump{L_{ijkl}(\bfy)\dfrac{\partial u^{(1)}_k}{\partial x_l}(\bfx,\bfy)}\widehat{N}_j-\vspace{0.2cm}\\
\hspace{2.75cm}\dfrac{\partial}{\partial x_q}\left[\widehat{L}_{ijkl}\left(\dfrac{\partial u_k}{\partial x_p}(\bfx)+\dfrac{\partial u^{(1)}_k}{\partial y_p}(\bfx,\bfy)\right)\widehat{I}_{pl}\right]\widehat{I}_{qj},\quad\bfy\in G
\end{array}\right. .\label{Eq-u2}
\end{align}
For any function $\bfu(\bfx)$ of choice, noting that $\bfu^{(1)}(\bfx,\bfy)$ is given by (\ref{Sol-u1}) in terms of $\bfu(\bfx)$, equations (\ref{Eq-u2}) are nothing more than a unit-cell problem for the function $\bfu^{(2)}(\bfx,\bfy)$, where once more $\bfx$ plays the role of a parameter.

Analogously to the classical context of elastostatics without residual stresses and interfacial forces (\cite{BLP11}, Chapter 2), equation (\ref{Eq-u2}) can be manipulated to yield the governing equation for the leading-order function (\ref{Sol-u0}) in the ansatz (\ref{ansatz}). Indeed, upon integrating equation (\ref{Eq-u2})$_1$ over $Y$, equation (\ref{Eq-u2})$_2$ over $G$, summing the two results together, then using the bulk divergence theorem
\begin{equation}
\displaystyle\int_{Y}\dfrac{\partial (\cdot)}{\partial y_j}\,{\rm d}\bfy=\displaystyle\int_{\partial Y}(\cdot) N_j\,{\rm d}\bfy+\displaystyle\int_{G} \jump{\cdot}\widehat{N}_{j}\,{\rm d}\bfy\label{Div-Bulk}
\end{equation}
and the interface divergence theorem
\begin{equation}
\displaystyle\int_{G}\dfrac{\partial (\cdot)}{\partial y_q}\widehat{I}_{qj}\,{\rm d}\bfy=\displaystyle\int_{G}\dfrac{\partial \widehat{N}_m}{\partial y_n}\widehat{I}_{mn}(\cdot) \widehat{N}_j\,{\rm d}\bfy, \label{Div-Interfaces}
\end{equation}
noting that $\widehat{L}_{ijkl}\widehat{N}_{j}=0$, and recognizing the identity $\widehat{L}_{ijkl}\widehat{I}_{qj}=\widehat{L}_{iqkl}$, it follows that
\begin{align*}
&\dfrac{\partial}{\partial x_j}\displaystyle\int_{Y} L_{ijkl}(\bfy)\left(\dfrac{\partial u_k}{\partial x_l}(\bfx)+\dfrac{\partial u^{(1)}_k}{\partial y_l}(\bfx,\bfy)\right){\rm d}\bfy+\nonumber\\
&\dfrac{\partial}{\partial x_j}\displaystyle\int_{G}\widehat{L}_{ijkl}\left(\dfrac{\partial u_k}{\partial x_p}(\bfx)+\dfrac{\partial u^{(1)}_k}{\partial y_p}(\bfx,\bfy)\right)\widehat{I}_{pl}{\rm d}\bfy=0 .
\end{align*}
Finally, making use of the representation (\ref{Sol-u1}) for $\bfu^{(1)}(\bfx,\bfy)$ in terms of the $Y$-periodic function $\omega_{kmn}(\bfy)$, it is a simple matter to deduce that this last relation can be rewritten in the form
\begin{align}
\dfrac{\partial}{\partial x_j}\left[\overline{L}_{ijkl}\dfrac{\partial u_{k}}{\partial x_l}(\bfx)\right]=0,\label{Hom-Eq-0}
\end{align}
where
\begin{align}
\overline{L}_{ijkl}=&\displaystyle\int_{Y} L_{ijmn}(\bfy)\left(\delta_{mk}\delta_{nl}+\dfrac{\partial \omega_{mkl}}{\partial y_n}(\bfy)\right){\rm d}\bfy+\nonumber\\
&\displaystyle\int_{G}\widehat{L}_{ijmn}\left(\delta_{mk}\widehat{I}_{nl}+\dfrac{\partial \omega_{mkl}}{\partial y_p}(\bfy)\widehat{I}_{pn}\right){\rm d}\bfy. \label{Leff-0}
\end{align}

Equation (\ref{Hom-Eq-0}) is the homogenized equation in ${\rm \Omega}$ that, together with the boundary condition $\bfu(\bfx)=\overline{\bfu}(\bfx)$ on $\partial{\rm \Omega}$, completely determines the macroscopic displacement field $\bfu(\bfx)$. The following remarks are in order:

\paragraph{i. Physical interpretation of the homogenized equation (\ref{Hom-Eq-0})} Equation (\ref{Hom-Eq-0}), together with the boundary condition $\bfu(\bfx)$ on $\partial{\rm \Omega}$, corresponds to the governing equation for the displacement field within a \emph{homogeneous} linear elastic solid, with constant effective modulus of elasticity $\overline{\bfL}$, undergoing small quasistatic deformations.

\paragraph{ii. Absence of a macroscopic residual stress} In spite of the fact that there is a local stress within the inclusions and an initial surface tension on the elastomer/inclusions interfaces, the homogenized equation (\ref{Hom-Eq-0}) is free of residual stresses. The reason behind this result is that the average of the local residual stress and initial surface tension cancel each other out. Precisely,
\begin{align}
-\displaystyle\int_{Y}\displaystyle\sum_{I=1}^{\texttt{N}}\theta_I(\bfy)\dfrac{(n-1)\widehat{\gamma}}{A_I}\bfI\,{\rm d}\bfy+\displaystyle\int_{G}\widehat{\gamma}\,\widehat{\bfI}\,{\rm d}\bfy=\textbf{0}.\label{zero-res-eff}
\end{align}

\paragraph{iii. The effective modulus of elasticity $\overline{\bfL}$} The effective modulus of elasticity (\ref{Leff-0}) that emerges in the homogenized equation (\ref{Hom-Eq-0}) is independent of the choice of the domain ${\rm \Omega}$ occupied by the filled elastomer and the boundary conditions on $\partial {\rm \Omega}$. It does depend, however, on the size of the inclusions, the residual hydrostatic stress that they are subjected to in the initial configuration, as well as on the elasticity of the interfaces and the surface tension that they are subjected to in the initial configuration.

\paragraph{iv. Symmetries of $\overline{\bfL}$} The effective modulus of elasticity (\ref{Leff-0}) satisfies the major and minor symmetries
\begin{align}
\overline{L}_{ijkl}=\overline{L}_{klij}\quad {\rm and}\quad \overline{L}_{ijkl}=\overline{L}_{jikl}=\overline{L}_{ijlk}\label{Symm_Leff}
\end{align}
of a conventional homogeneous elastic solid, this in spite of the fact that the local moduli of elasticity $\bfL(\bfy)$ and $\widehat{\bfL}$ for the bulk and the interfaces do \emph{not} possess minor symmetries.

The major symmetry $\overline{L}_{ijkl}=\overline{L}_{klij}$ is a direct consequence of the fact that the local moduli $\bfL(\bfy)$ and $\widehat{\bfL}$ themselves possess major symmetry. To see this, making use of the bulk (\ref{Div-Bulk}) and interface (\ref{Div-Interfaces}) divergence theorems, as well as of the definition (\ref{Eq-omega}) for the $Y$-periodic corrector function $\boldsymbol{\omega}(\bfy)$, first note that
\begin{align*}
&\displaystyle\int_{Y}\dfrac{\partial \omega_{mij}}{\partial y_n}(\bfy)L_{mnpq}(\bfy)\left(\delta_{pk}\delta_{ql}+\dfrac{\partial \omega_{pkl}}{\partial y_q}(\bfy)\right){\rm d}\bfy+\displaystyle\int_{G}\dfrac{\partial \omega_{mij}}{\partial y_r}(\bfy)\widehat{I}_{rn}\widehat{L}_{mnpq}\times\\
&\left(\delta_{pk}\widehat{I}_{ql}+
\dfrac{\partial \omega_{pkl}}{\partial y_s}(\bfy)\widehat{I}_{sq}\right){\rm d}\bfy=\displaystyle\int_{Y}\dfrac{\partial}{\partial y_n}\left[\omega_{mij}(\bfy)L_{mnpq}(\bfy)\left(\delta_{pk}\delta_{ql}+\dfrac{\partial \omega_{pkl}}{\partial y_q}(\bfy)\right)\right]{\rm d}\bfy-\\
&\displaystyle\int_{Y}\omega_{mij}(\bfy)\dfrac{\partial}{\partial y_n}\left[L_{mnpq}(\bfy)\left(\delta_{pk}\delta_{ql}+\dfrac{\partial \omega_{pkl}}{\partial y_q}(\bfy)\right)\right]{\rm d}\bfy+\displaystyle\int_{G}\dfrac{\partial}{\partial y_r}\left[ \omega_{mij}(\bfy)\widehat{L}_{mnpq}\times\right.\\
&\left.\left(\delta_{pk}\widehat{I}_{ql}+
\dfrac{\partial \omega_{pkl}}{\partial y_s}(\bfy)\widehat{I}_{sq}\right)\right]\widehat{I}_{rn}{\rm d}\bfy-\displaystyle\int_{G}\omega_{mij}(\bfy)\dfrac{\partial }{\partial y_r}\left[\widehat{L}_{mnpq}\left(\delta_{pk}\widehat{I}_{ql}+
\dfrac{\partial \omega_{pkl}}{\partial y_s}(\bfy)\widehat{I}_{sq}\right)\right]\times\\
&\widehat{I}_{rn}{\rm d}\bfy=\displaystyle\int_{G}\jump{\omega_{mij}(\bfy)L_{mnpq}(\bfy)\left(\delta_{pk}\delta_{ql}+\dfrac{\partial \omega_{pkl}}{\partial y_q}(\bfy)\right)}\widehat{N}_n{\rm d}\bfy-\\
&\displaystyle\int_{G}\omega_{mij}(\bfy)\jump{L_{mnpq}(\bfy)\left(\delta_{pk}\delta_{ql}+\dfrac{\partial \omega_{pkl}}{\partial y_q}(\bfy)\right)}\widehat{N}_{n}{\rm d}\bfy=0.
\end{align*}
With this result at hand, it is a simple matter to verify that the formula (\ref{Leff-0}) can be rewritten in the equivalent form
\begin{align*}
\overline{L}_{ijkl}=&\displaystyle\int_{Y}\left(\delta_{mi}\delta_{nj}+\dfrac{\partial \omega_{mij}}{\partial y_n}(\bfy)\right)L_{mnpq}(\bfy)\left(\delta_{pk}\delta_{ql}+\dfrac{\partial \omega_{pkl}}{\partial y_q}(\bfy)\right){\rm d}\bfy+\\
&\displaystyle\int_{G}\left(\delta_{mi}\widehat{I}_{nj}+\dfrac{\partial \omega_{mij}}{\partial y_r}(\bfy)\widehat{I}_{rn}\right)\widehat{L}_{mnpq}\left(\delta_{pk}\widehat{I}_{ql}+\dfrac{\partial \omega_{pkl}}{\partial y_s}(\bfy)\widehat{I}_{sq}\right){\rm d}\bfy, 
\end{align*}
from which it is trivial to establish that $\overline{L}_{ijkl}=\overline{L}_{klij}$ since $L_{mnpq}(\bfy)=L_{pqmn}(\bfy)$ and $\widehat{L}_{mnpq}=\widehat{L}_{pqmn}$.

On the other hand, the minor symmetries $\overline{L}_{ijkl}=\overline{L}_{jikl}$ and $\overline{L}_{ijkl}=\overline{L}_{ijlk}$ are a direct consequence of the absence of a macroscopic residual stress (\ref{zero-res-eff}) and the macroscopic major symmetry (\ref{Symm_Leff})$_1$ of $\overline{\bfL}$ . To see this, first note that
\begin{align*}
&-\displaystyle\int_{Y}\displaystyle\sum_{I=1}^{\texttt{N}}\theta_I(\bfy)\dfrac{(n-1)\widehat{\gamma}}{A_I}\left(\delta_{il}\delta_{jk}+\dfrac{\partial \omega_{jkl}}{\partial y_i}(\bfy)\right)\,{\rm d}\bfy+\displaystyle\int_{G}\widehat{\gamma}\left(\delta_{jk}\widehat{I}_{il}+\dfrac{\partial \omega_{jkl}}{\partial y_p}(\bfy)\widehat{I}_{ip}\right)\,{\rm d}\bfy=\\
&-\displaystyle\int_{Y}\displaystyle\sum_{I=1}^{\texttt{N}}\theta_I(\bfy)\dfrac{(n-1)\widehat{\gamma}}{A_I}\dfrac{\partial \omega_{jkl}}{\partial y_i}(\bfy)\,{\rm d}\bfy+\displaystyle\int_{G}\widehat{\gamma}\dfrac{\partial \omega_{jkl}}{\partial y_p}(\bfy)\widehat{I}_{ip}\,{\rm d}\bfy=-\displaystyle\int_{Y}\dfrac{\partial}{\partial y_i}\left[\displaystyle\sum_{I=1}^{\texttt{N}}\theta_I(\bfy)\times\right.\\
&\left.\dfrac{(n-1)\widehat{\gamma}}{A_I}\omega_{jkl}(\bfy)\right]{\rm d}\bfy+\displaystyle\int_{Y}\dfrac{\partial}{\partial y_i}\left[\displaystyle\sum_{I=1}^{\texttt{N}}\theta_I(\bfy)\dfrac{(n-1)\widehat{\gamma}}{A_I}\right]\omega_{jkl}(\bfy){\rm d}\bfy+\displaystyle\int_{G}\widehat{\gamma}\dfrac{\partial\widehat{N}_m}{\partial y_n}\widehat{I}_{mn}\widehat{N}_i\times\\
&\omega_{jkl}(\bfy)\,{\rm d}\bfy=\displaystyle\sum_{I=1}^{\texttt{N}}\left(-\displaystyle\int_{G_I}\dfrac{(n-1)\widehat{\gamma}}{A_I}\omega_{jkl}(\bfy)\widehat{N}_i{\rm d}\bfy+\displaystyle\int_{G_I}\dfrac{(n-1)\widehat{\gamma}}{A_I}\omega_{jkl}(\bfy)\widehat{N}_i{\rm d}\bfy\right)=0
,
\end{align*}
where $G_I$ denotes the interface of the $I$th inclusion and where use has been made of relation  (\ref{zero-res-eff}), the bulk (\ref{Div-Bulk}) and interface (\ref{Div-Interfaces}) divergence theorems, as well as of the $Y$-periodicity of the corrector function $\boldsymbol{\omega}(\bfy)$. In view of this last result, it is straightforward to show that the formula (\ref{Leff-0}) can also be rewritten as
\begin{align*}
\overline{L}_{ijkl}=&\displaystyle\int_{Y}\left(L_{ijmn}(\bfy)-\displaystyle\sum_{I=1}^{\texttt{N}}\theta_I(\bfy)\dfrac{(n-1)\widehat{\gamma}}{A_I}\delta_{in}\delta_{jm}\right)\left(\delta_{mk}\delta_{nl}+\dfrac{\partial \omega_{mkl}}{\partial y_n}(\bfy)\right)\,{\rm d}\bfy+\nonumber\\
&\displaystyle\int_{G}\left(\widehat{L}_{ijmn}+\widehat{\gamma}\delta_{jm}\widehat{I}_{in}\right)\left(\delta_{mk}\widehat{I}_{nl}+\dfrac{\partial \omega_{mkl}}{\partial y_p}(\bfy)\widehat{I}_{pn}\right)\,{\rm d}\bfy, 
\end{align*}
from which it is trivial to establish that $\overline{L}_{ijkl}=\overline{L}_{jikl}$ since the combinations $L_{ijmn}(\bfy)$ $-(\sum_{I=1}^{\texttt{N}}\theta_I(\bfy)(n-1)\widehat{\gamma}/A_I) \delta_{in}\delta_{jm}$ and $\widehat{L}_{ijmn}+\widehat{\gamma}\delta_{jm}\widehat{I}_{in}$ possess minor symmetries. Minor symmetries in the last two indices $\overline{L}_{ijkl}=\overline{L}_{ijlk}$ can be established by exploiting the major symmetry $\overline{L}_{ijkl}=\overline{L}_{klij}$ and then following the same steps as above.

\paragraph{v. Positive definiteness of $\overline{\bfL}$} Physically, the expectation is that the effective modulus of elasticity (\ref{Leff-0}) be positive definite. However, given that the local moduli of elasticity $\bfL(\bfy)$ and $\widehat{\bfL}$ for the bulk and the interfaces are \emph{not} positive definite in general, the standard argument (\cite{BLP11}, Section 2.3 of Chapter 1) to prove so does \emph{not} apply here. This difficulty is intimately related to the difficulty of proving existence of solution for the boundary-value problem (\ref{BVP-1}) noted at the end of the preceding section. We shall address both of these issues in a separate contribution.

\paragraph{vi. Computation of $\overline{\bfL}$} The computation of the effective modulus of elasticity (\ref{Leff-0}) amounts to solving the unit-cell problem (\ref{Eq-omega}) for the corrector $\boldsymbol{\omega}(\bfy)$. In general, this can only be accomplished numerically. Ghosh and Lopez-Pamies \cite{GLP22} have recently put forth a finite-element (FE) scheme to generate numerical solutions for such classes of boundary-value problems. In the next section, by way of an example, we make use of that scheme to generate solutions for the effective modulus of elasticity of isotropic suspensions of incompressible liquid $2$-spherical inclusions of monodisperse size embedded in an isotropic incompressible elastomer.

\paragraph{vii. Strain and stress macro-variables} A quick glance at the homogenized equation (\ref{Hom-Eq-0}) suffices to identify
\begin{equation}
H_{ij}(\bfx)= \dfrac{\partial u_i}{\partial x_j}(\bfx)\label{H-macro}
\end{equation}
as the macroscopic displacement gradient field and
\begin{equation}
S_{ij}(\bfx)= \overline{L}_{ijkl}\dfrac{\partial u_k}{\partial x_l}(\bfx)\label{S-macro}
\end{equation}
as the macroscopic stress measure that describe the constitutive response of the resulting effective elastic solid in the homogenization limit.

By virtue of the minor symmetries (\ref{Symm_Leff})$_2$ of the effective modulus of elasticity $\overline{\bfL}$, remark that the constitutive relation between (\ref{H-macro}) and (\ref{S-macro}) can be written in the classical stress-strain form
\begin{equation}
S_{ij}(\bfx)= \overline{L}_{ijkl}E_{kl}(\bfx),\quad E_{ij}(\bfx):=\dfrac{1}{2}\left(H_{ij}(\bfx)+H_{ji}(\bfx)\right).\label{S-E-macro}
\end{equation}

The macro-variable (\ref{H-macro}) happens to be identical to the one that arises in the classical context of elastostatics without residual stresses and interfacial forces (\cite{BLP11}, Chapter 2). Precisely,
\begin{align*}
H_{ij}(\bfx)=& \displaystyle\int_{Y}\left(\dfrac{\partial u_i}{\partial x_j}(\bfx)+\dfrac{\partial u^{(1)}_i}{\partial y_j}(\bfx,\bfy)\right){\rm d}\bfy\nonumber\\
=&\dfrac{\partial u_i}{\partial x_j}(\bfx)+\displaystyle\int_{\partial Y} u^{(1)}_i(\bfx,\bfy) N_j {\rm d}\bfy+\displaystyle\int_{G} \jump{u^{(1)}_i(\bfx,\bfy)} \widehat{N}_j {\rm d}\bfy\nonumber\\
=&\dfrac{\partial u_i}{\partial x_j}(\bfx).
\end{align*}

By contrast, the macro-variable (\ref{S-macro}) is \emph{not} in accord with the classical result. Instead, relation (\ref{S-macro}) corresponds to the average over the unit cell $Y$ of the local stress in the bulk \emph{plus} the average over the interfaces $G$ of the local interface stress. Precisely,
\begin{align*}
S_{ij}(\bfx)=& \displaystyle\int_{Y}L_{ijkl}(\bfy)\left(\dfrac{\partial u_k}{\partial x_l}(\bfx)+\dfrac{\partial u^{(1)}_k}{\partial y_l}(\bfx,\bfy)\right){\rm d}\bfy+\nonumber\\
&\displaystyle\int_{G}\widehat{L}_{ijkl}\left(\dfrac{\partial u_k}{\partial x_p}(\bfx)+\dfrac{\partial u^{(1)}_k}{\partial y_p}(\bfx,\bfy)\right)\widehat{I}_{pl}{\rm d}\bfy\nonumber\\
=&\overline{L}_{ijkl}\dfrac{\partial u_k}{\partial x_l}(\bfx).
\end{align*}
A similar result emerges in the homogenization of elastic dielectric composites containing space charges \cite{LLP17,FGLP21}.

\paragraph{viii. Effective stored-energy function} By virtue of the major symmetry (\ref{Symm_Leff})$_1$ of the effective modulus of elasticity $\overline{\bfL}$, the macroscopic constitutive relation (\ref{S-E-macro}) is a hyperelastic one. That is, there is an effective stored-energy function, $\overline{W}(\textbf{E})$ say, whose derivative with respect to the macroscopic strain $\textbf{E}$ yields the macroscopic stress $\bfS$.

Precisely, making use of the bulk (\ref{Div-Bulk}) and interface (\ref{Div-Interfaces}) divergence theorems, together with the representation (\ref{Sol-u1}) for $\bfu^{(1)}(\bfx,\bfy)$ and the definition (\ref{Eq-omega}) for the $Y$-periodic corrector function $\boldsymbol{\omega}(\bfy)$, it is not difficult to deduce that
\begin{equation*}
S_{ij}=\dfrac{\partial \overline{W}}{\partial E_{ij}}(\textbf{E}),
\end{equation*}
where
\begin{align*}
\overline{W}(\textbf{E})=&\dfrac{1}{2}\displaystyle\int_{Y}\left(\dfrac{\partial u_i}{\partial x_j}(\bfx)+\dfrac{\partial u^{(1)}_i}{\partial y_j}(\bfx,\bfy)\right)L_{ijkl}(\bfy)\left(\dfrac{\partial u_k}{\partial x_l}(\bfx)+\dfrac{\partial u^{(1)}_k}{\partial y_l}(\bfx,\bfy)\right)\,{\rm d}\bfy+\nonumber\\
&\dfrac{1}{2}\displaystyle\int_{G}\left(\dfrac{\partial u_i}{\partial x_p}(\bfx)+\dfrac{\partial u^{(1)}_i}{\partial y_p}(\bfx,\bfy)\right)\widehat{I}_{pj}\widehat{L}_{ijkl}\left(\dfrac{\partial u_k}{\partial x_q}(\bfx)+\dfrac{\partial u^{(1)}_k}{\partial y_q}(\bfx,\bfy)\right)\widehat{I}_{ql}\,{\rm d}\bfy \nonumber\\
=&\dfrac{1}{2} E_{ij} \overline{L}_{ijkl} E_{kl}.
\end{align*}

\section{The homogenized behavior of isotropic suspensions of monodisperse $2$-spherical inclusions}\label{Sec: Application}

In this final section, for demonstration purposes, we present numerical results for the effective modulus of elasticity $\overline{\bfL}$ of a basic class of elastomers filled with liquid inclusions, that of isotropic suspensions of $2$-spherical inclusions of monodisperse size,
\begin{equation*}
A_I=A\quad I=1,...,\texttt{N}
\end{equation*}
made of an incompressible liquid,
\begin{equation*}
\Lambda^{(\texttt{i})}=+\infty,
\end{equation*}
embedded in an isotropic incompressible elastomer,
\begin{equation*}
\bfL^{(\texttt{m})}=2\mu^{(\texttt{m})}\Ktan+\infty \Jtan,
\end{equation*}
wherein the interfaces only feature a constant surface tension $\widehat{\gamma}$ and hence the interface Lam\'e constants
\begin{equation*}
\widehat{\mu}=\widehat{\Lambda}=0.
\end{equation*}
For this fundamental class of filled elastomers, remark that there is a sole dimensionless material constant that describes the constitutive behavior, the so-called elasto-capillary number
\begin{equation*}
eCa:=\dfrac{\widehat{\gamma}}{2\mu^{(\texttt{m})}A}.
\end{equation*}
Physically, $eCa$ is a measure of interface stiffness $\widehat{\gamma}/2A$ relative to bulk stiffness $\mu^{(\texttt{m})}$ \cite{Andreottietal16,Bicoetal18}.

\subsection{Construction of the unit cells $Y$}\label{Sec:Microstructures}

Prior to the presentation of the results for $\overline{\bfL}$ \emph{per se} in Subsection \ref{Sec:Results}, we begin by outlining the process by which we constructed the unit cells $Y$.

We follow in the footstep of a well-settled approach \cite{Gusev97,LPGD13} and approximate the aforementioned class of \emph{isotropic} filled elastomers as infinite media made of the periodic repetition of unit cells $Y$ that contain random distributions of a sufficiently large number $\texttt{N}$ of inclusions. A critical point in this approach is to determine what that sufficiently large number $\texttt{N}$ is so that the resulting homogenized constitutive behaviors are indeed isotropic to a high enough degree of accuracy.

In order to cover a large range of inclusion concentrations (that is, in the present context of $n=2$ space dimensions, area fractions of inclusions)
\begin{equation*}
c:=\displaystyle\int_{Y}\theta(\bfy){\rm d}\bfy,
\end{equation*}
we make use of the algorithm introduced by Lubachevsky and Stillinger \cite{LS90}. Roughly speaking, the idea behind this algorithm is to randomly seed at once in the unit cell $Y$ the desired total number $\texttt{N}$ of inclusions as points endowed with random velocities and a uniform radial growth rate. As the points move and grow into $2$-spheres, their collision with one another are described by conservation of momentum, while their crossings through the boundaries of the unit cell are described by periodicity. When the desired concentration $c$ is reached, the algorithm is stopped.

Although the algorithm allows to generate microstructures spanning the full range of concentrations --- from the dilute limit $c\searrow 0$ to the percolation threshold $c\nearrow c_p\approx 0.90$ \cite{LS90} --- we do not wish to deal with the computational challenges of extremely packed microstructures and restrict our attention here to the range $c\in[0,0.50]$; the full range of concentrations will be considered in a companion work \cite{GLLP23}. Specifically, the construction process that we carried out is as follows.

In the footstep of \cite{LFLP22,LLP22}, we started by generating a total of 10,800 realizations of unit cells $Y=(0,1)^2$ containing 30, 60, 120, 240, 480, 960 randomly distributed inclusions with six different concentrations $c=0.05, 0.10, 0.20, 0.30, 0.40, 0.50$ and three different minimum inter-inclusion distances $d=0.01 A, 0.02 A, 0.05A$. For each realization, we computed the two-point correlation function $P_2(\bfy)=\int_{Y}\theta(\bfy^\prime)\theta(\bfy+\bfy^\prime)\,{\rm d}\bfy^\prime$. As a first assessment of deviation from exact geometric isotropy (which is only achieved in the limit of infinitely many inclusions), we then computed the deviation of $P_2(\bfy)$ from its isotropic projection $I_2(|\bfy|)={1}/(2\pi)\int_0^{2\pi}P_2(|\bfy|\cos\phi \textbf{e}_1+|\bfy|\sin\phi \textbf{e}_2)\,{\rm d}\phi$ onto the space of functions that depend on $\bfy$ only through its magnitude $|\bfy|$; recall that $\{\textbf{e}_1,\textbf{e}_2\}$ stand for the principal axes of the unit cell $Y$. Realizations that did not satisfy the condition
\begin{equation}
\dfrac{||P_2(\bfy)-I_2(|\bfy|)||_1}{||I_2(|\bfy|)||_1}\le  10^{-2}\label{P2}
\end{equation}
were discarded as not sufficiently isotropic. This filtering process reduced the initial set of 10,800 realizations to just a set of 90 potentially acceptable realizations, five for each of the six concentrations $c=0.05, 0.10, 0.20, 0.30, 0.40, 0.50$ and the three minimum inter-inclusion distances $d=0.01 A, 0.02 A, 0.05A$.

\begin{figure}[t!]
   \centering \includegraphics[width=4.6in]{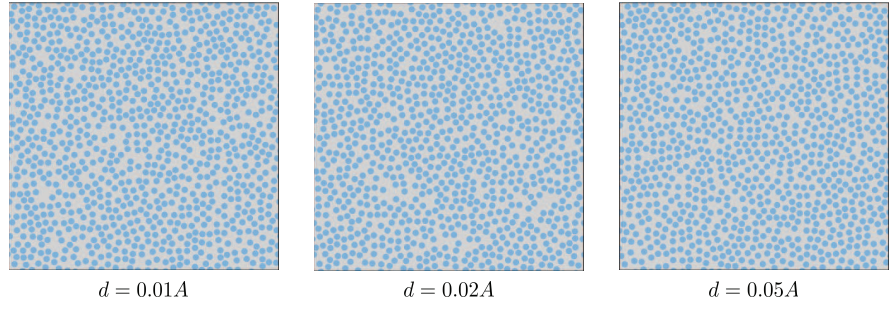}
   \caption{{\small Representative unit cells $Y$ containing random distributions of $\overline{\bfL}_{iso}=960$ $2$-spherical inclusions of monodisperse radius $A$ at concentration $c=0.50$ and minimum distances $d=0.01 A, 0.02A,$ and $0.05 A$ between the inclusions.}}\label{Fig2}
\end{figure}

Thanks to its pure geometric nature, the criterion (\ref{P2}) provides a computationally inexpensive tool to weed out microstructures that are unlikely to lead to isotropic constitutive behaviors. However, microstructures that do satisfy (\ref{P2}) need not exhibit isotropic constitutive behaviors. To conclusively establish whether a given realization with a \emph{finite} number $\texttt{N}$ of inclusions does indeed exhibit isotropic constitutive behavior to within the desired accuracy, one needs to compute its effective modulus of elasticity $\overline{\bfL}$ in its entirety and then quantify its deviation from exact constitutive isotropy. Accordingly, for each of the 90 potentially acceptable realizations and each of the three elasto-capillary numbers $eCa=0.20,1,5$ that we considered in this study, we generated numerical solutions for the entire $\overline{\bfL}$ via the ($n=2$ version of the) FE scheme put forth in \cite{GLP22} and then computed its isotropic deviatoric projection
\begin{equation}
\overline{\bfL}_{iso}=2\,\overline{\mu}\,\Ktan,\qquad \overline{\mu}:=\dfrac{1}{4}\,\Ktan\cdot\overline{\bfL}=\dfrac{1}{4}\,\mathcal{K}_{ijkl}\overline{L}_{ijkl} ,\label{mueff}
\end{equation}
which serves to define the effective shear modulus $\overline{\mu}$ of the filled elastomer at hand. Realizations that did not satisfy the stringent threshold
\begin{equation}
\dfrac{||\Ktan\,\overline{\bfL}\,\Ktan-\overline{\bfL}_{iso}||_{\infty}}{||\Ktan\,\overline{\bfL}\,\Ktan||_{\infty}}\leq 0.02\label{L-iso}
\end{equation}
were discarded as not sufficiently isotropic. Those that did satisfy (\ref{L-iso}) are the ones for which we present results below. Importantly, the maximum difference between any two such realizations with the same inclusion concentration $c$ and the same minimum inter-inclusion distance $d$ was less than $2\%$, and hence, as expected \cite{Papanicolaou81}, they exhibited practically the same homogenized behavior. By way of an example, Fig. \ref{Fig2} shows three representative unit cells $Y$ containing a total of $\overline{\bfL}_{iso}=960$ inclusions at concentration $c=0.50$ and minimum inter-inclusion distances $d=0.01 A, 0.02A,$ and $0.05 A$ that satisfy conditions (\ref{P2}) and (\ref{L-iso}).

%
%

\subsection{Results}\label{Sec:Results}

%
\begin{figure}[b!]
\centering
\begin{subfigure}{0.48\textwidth}
\centering
\includegraphics[width=0.97\linewidth]{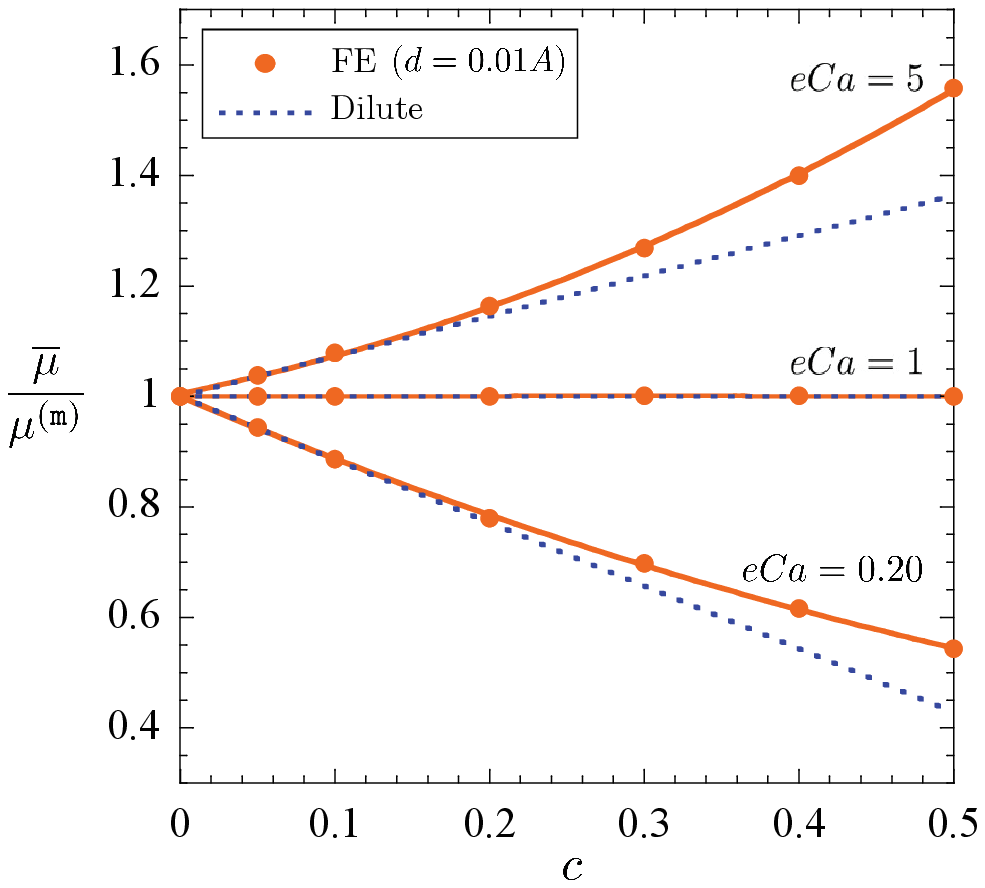}
\caption*{(a)}
\end{subfigure}\hspace{2mm}%
\begin{subfigure}{0.48\textwidth}
\centering
\includegraphics[width=0.97\linewidth]{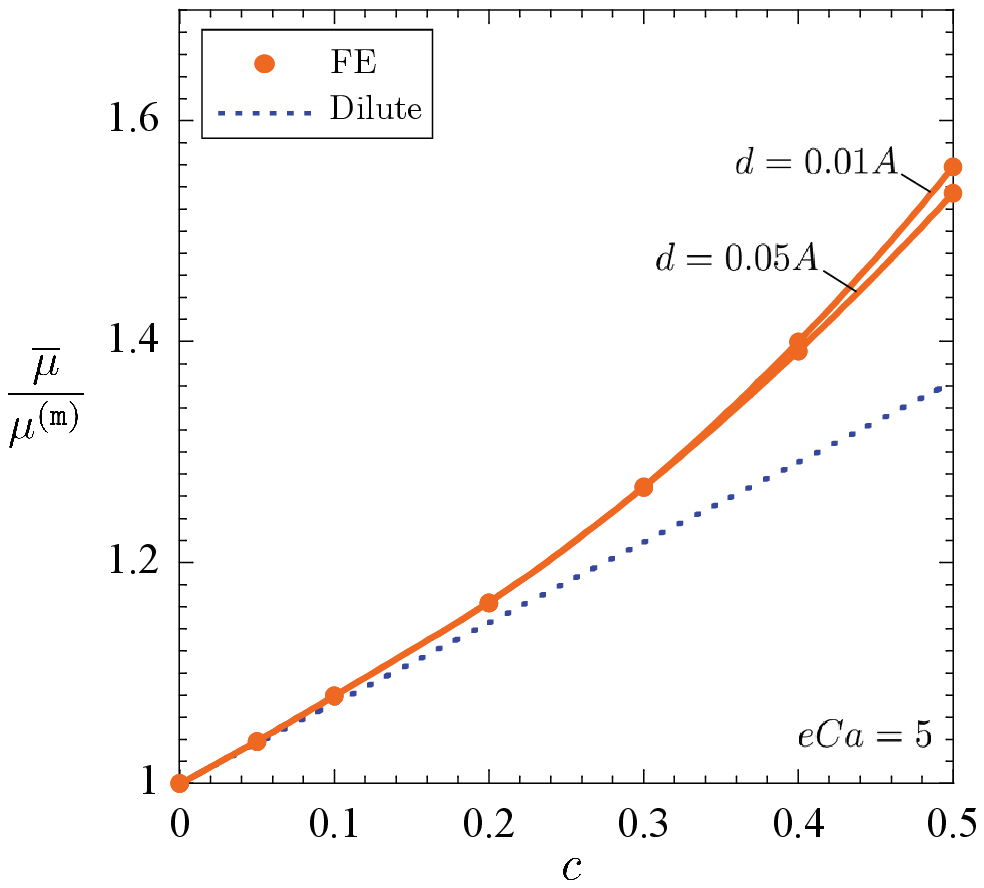}
\caption*{(b)}
\end{subfigure}
\caption{{\small The effective shear modulus $\overline{\mu}$, normalized by the shear modulus of the underlying elastomeric matrix $\mu^{(\texttt{m})}$, for isotropic suspensions of monodisperse $2$-spherical liquid inclusions spanning a range of concentrations $c$ of inclusions and minimum inter-inclusion distances $d$. (a) $\overline{\mu}/\mu^{(\texttt{m})}$ as a function of $c$ for $d=0.01 A$ and three values of the elasto-capillary number $eCa$. (b) $\overline{\mu}/\mu^{(\texttt{m})}$ as a function of $c$ for $d=0.01 A, 0.05 A$ and elasto-capillary number $eCa=5$. For direct comparison, the plots include the asymptotic result (\ref{mueff-dilute}) for the corresponding dilute suspension.}}\label{Fig3}
\end{figure}
%

Figure \ref{Fig3} presents the FE solutions obtained for the effective shear modulus $\overline{\mu}$, as defined in (\ref{mueff}), of the isotropic suspensions described above. While Fig. \ref{Fig3}(a) shows the effective shear modulus $\overline{\mu}$, normalized by the shear modulus of the elastomeric matrix $\mu^{(\texttt{m})}$, for minimum inter-inclusion distance $d=0.01A$ and elasto-capillary numbers $eCa=0.20,1,5$ as a function of the concentration $c$ of inclusions, Fig. \ref{Fig3}(b) shows $\overline{\mu}/\mu_{\texttt{m}}$ as a function of $c$ for $d=0.01A, 0.05 A$ and elasto-capillary number $eCa=5$. For completeness, all plots include the asymptotic result
\begin{align}
\overline{\mu}=\overline{\mu}^{\,{\rm dil}}+O(c^2),\qquad\overline{\mu}^{\,{\rm dil}}=&\mu^{(\texttt{m})}+\dfrac{(2+n)(eCa-1)}{n+(2+n)\, eCa}\,\mu^{(\texttt{m})}\, c \nonumber\\
=&\mu^{(\texttt{m})}+\dfrac{2(eCa-1)}{1+2\, eCa}\,\mu^{(\texttt{m})}\, c\label{mueff-dilute}
\end{align}
for the effective shear modulus of a dilute suspension; see, e.g., Appendix D in \cite{GLP22} for a derivation of this result in space dimension $n=3$. To be precise, the result (\ref{mueff-dilute}) corresponds to the response of an infinitely large elastomer domain that contains a \emph{single} liquid inclusion. In other words, the result (\ref{mueff-dilute}) is an extension of the classical result of Eshelby \cite{Eshelby57} to account for the presence of surface tension at the matrix/inclusion interface.

Three observations are immediate from Fig. \ref{Fig3}. First, irrespectively of the concentration $c$ of inclusions, $\overline{\mu}<\mu^{(\texttt{m})}$ for $eCa=0.20<1$, $\overline{\mu}=\mu^{(\texttt{m})}$ for $eCa=1$, and $\overline{\mu}>\mu^{(\texttt{m})}$ for $eCa=5>1$. That is, while the presence of liquid inclusions leads to the \emph{softening} of the material when $eCa<1$, it leads to \emph{stiffening} when $eCa>1$. The transition from softening to stiffening occurs precisely at $eCa=1$, when, rather interestingly, the presence of liquid inclusions goes unnoticed in the homogenized response. This behavior can be readily understood by recognizing that liquid inclusions with ``small'' interface stiffness $\widehat{\gamma}/2A$ pose little resistance to deformation and hence lead to the softening of the homogenized response. By contrast, inclusions with ``large'' interface stiffness $\widehat{\gamma}/2A$ pose significant resistance to deformation, behave effectively as stiff inclusions, and hence lead to the stiffening of the homogenized response. Second, both the softening and the stiffening can be very significant even at moderate values of $c$ and $eCa$. At $c=0.5$, for instance, we see from Fig. \ref{Fig3}(a) that $\overline{\mu}=0.54\mu^{(\texttt{m})}$ for $eCa=0.20$ and $\overline{\mu}=1.56\mu^{(\texttt{m})}$ for $eCa=5$. Finally, the minimum inter-inclusion distance $d$ remains inconsequential from the dilute limit $c\searrow 0$ up to approximately $c\approx 0.40$. For larger concentrations of inclusions, as expected on physical grounds \cite{LLP22}, suspensions with different minimum inter-inclusion distances $d$ can exhibit sizably different responses, more so the larger the concentration.

\section*{Acknowledgements}

\noindent Support for this work by the National Science Foundation through the Grant DMREF--1922371 is gratefully acknowledged. V.L. would also like to acknowledge support through the computational resources and staff contributions provided for the Quest high performance computing facility at Northwestern University which is jointly supported by the Office of the Provost, the Office for Research, and Northwestern University Information Technology.

\bibliographystyle{unsrtnat}
\bibliography{references}

\begin{thebibliography}{26}
\providecommand{\natexlab}[1]{#1}
\providecommand{\url}[1]{\texttt{#1}}
\expandafter\ifx\csname urlstyle\endcsname\relax
  \providecommand{\doi}[1]{doi: #1}\else
  \providecommand{\doi}{doi: \begingroup \urlstyle{rm}\Url}\fi

\bibitem[Lopez-Pamies(2014)]{LP14}
O.~Lopez-Pamies.
\newblock Elastic dielectric composites: {T}heory and application to
  particle-filled ideal dielectrics.
\newblock \emph{J. Mech. Phys. Solids}, 64:\penalty0 61--82, 2014.

\bibitem[Style et~al.(2015{\natexlab{a}})Style, Boltyanskiy, Benjamin, Jensen,
  Foote, Wettlaufer, and Dufresne]{Syleetal15}
R.~W. Style, R.~Boltyanskiy, A.~Benjamin, K.~E. Jensen, H.~P. Foote, J.~S.
  Wettlaufer, and E.~R. Dufresne.
\newblock Stiffening solids with liquid inclusions.
\newblock \emph{Nature Physics}, 11:\penalty0 82--87, 2015{\natexlab{a}}.

\bibitem[Bartlett et~al.(2017)Bartlett, Kazem, Powell-Palm, Huang, Sun, Malen,
  and Majidi]{Bartlettetal2017}
M.~D. Bartlett, N.~Kazem, M.~J. Powell-Palm, X.~Huang, W.~Sun, J.~A. Malen, and
  C.~Majidi.
\newblock High thermal conductivity in soft elastomers with elongated liquid
  metal inclusions.
\newblock \emph{Proceedings of the National Academy of Sciences}, 114:\penalty0
  2143--2148, 2017.

\bibitem[Lef\`evre et~al.(2017)Lef\`evre, Danas, and Lopez-Pamies]{LDLP17}
V.~Lef\`evre, K.~Danas, and O.~Lopez-Pamies.
\newblock A general result for the magnetoelastic response of isotropic
  suspensions of iron and ferrofluid particles in rubber, with applications to
  spherical and cylindrical specimens.
\newblock \emph{J. Mech. Phys. Solids}, 107:\penalty0 343--364, 2017.

\bibitem[Lef\`evre et~al.(2019)Lef\`evre, Garnica, and Lopez-Pamies]{LGLP19}
V.~Lef\`evre, A.~Garnica, and O.~Lopez-Pamies.
\newblock A {WENO} finite-difference scheme for a new class of
  {H}amilton-{J}acobi equations in nonlinear solid mechanics.
\newblock \emph{Computer Methods in Applied Mechanics and Engineering},
  349:\penalty0 17--44, 2019.

\bibitem[Yun et~al.(2019)Yun, Tang, Sun, Yuan, Zhao, Deng, Yan, Du, Dickey, and
  Li]{Yunetal19}
G.~Yun, S.~Y. Tang, S.~Sun, D.~Yuan, Q.~Zhao, L.~Deng, S.~Yan, H.~Du, M.~D.
  Dickey, and W.~Li.
\newblock Liquid metal-filled magnetorheological elastomer with positive
  piezoconductivity.
\newblock \emph{Nature Communications}, 10:\penalty0 1300, 2019.

\bibitem[Ghosh and Lopez-Pamies(2022)]{GLP22}
K.~Ghosh and O.~Lopez-Pamies.
\newblock Elastomers filled with liquid inclusions: Theory, numerical
  implementation, and some basic results.
\newblock \emph{J. Mech. Phys. Solids}, 166:\penalty0 104930, 2022.

\bibitem[Coxeter(1973)]{Coxeter73}
H.~S.~M. Coxeter.
\newblock \emph{Regular Polytopes}.
\newblock Dover, Mineola, NY, 1973.

\bibitem[do~{C}armo(2016)]{Carmo16}
M.~P. do~{C}armo.
\newblock \emph{Differential {G}eometry of {C}urves and {S}urfaces}.
\newblock Dover, Mineola, 2016.

\bibitem[Gurtin et~al.(1998)Gurtin, Weissm\"uller, and Larch\'e]{GWL98}
M.~E. Gurtin, J.~Weissm\"uller, and F.~Larch\'e.
\newblock A general theory of curved deformable interfaces in solids at
  equilibrium.
\newblock \emph{Philosophical Magazine A}, 78:\penalty0 1093--1109, 1998.

\bibitem[Sharma et~al.(2003)Sharma, Ganti, and Bhate]{Sharmaetal03}
P.~Sharma, S.~Ganti, and N.~Bhate.
\newblock Effect of surfaces on the size-dependent elastic state of
  nano-inhomogeneities.
\newblock \emph{Appl. Phy. Letters}, 82:\penalty0 535--537, 2003.

\bibitem[Style et~al.(2015{\natexlab{b}})Style, Wettlaufer, and
  Dufresne]{Syleetal15b}
R.~W. Style, J.~S. Wettlaufer, and E.~R. Dufresne.
\newblock Surface tension and the mechanics of liquid inclusions in compliant
  solids.
\newblock \emph{Soft Matter}, 11:\penalty0 672--679, 2015{\natexlab{b}}.

\bibitem[Sanchez-Palencia(1980)]{SP80}
E.~Sanchez-Palencia.
\newblock \emph{Nonhomogeneous {M}edia and {V}ibration Theory}, volume 127 of
  \emph{Lecture Notes in Physics}.
\newblock Springer-Verlag, New York, 1980.

\bibitem[Bensoussan et~al.(2011)Bensoussan, Lions, and Papanicolau]{BLP11}
A.~Bensoussan, J.~L. Lions, and G.~Papanicolau.
\newblock \emph{Asymptotic Analysis for Periodic Structures}.
\newblock AMS, Chelsea, Providence, 2011.

\bibitem[Lef\`evre and Lopez-Pamies(2017)]{LLP17}
V.~Lef\`evre and O.~Lopez-Pamies.
\newblock Homogenization of elastic dielectric composites with rapidly
  oscillating passive and active source terms.
\newblock \emph{SIAM Journal on Applied Mathematics}, 77:\penalty0 1962--1988,
  2017.

\bibitem[Francfort et~al.(2021)Francfort, Gloria, and Lopez-Pamies]{FGLP21}
G.A. Francfort, A.~Gloria, and O.~Lopez-Pamies.
\newblock Enhancement of elasto-dielectrics by homogenization of active
  charges.
\newblock \emph{Journal de Math\'ematiques Pures et Appliqu\'ees},
  156:\penalty0 392--419, 2021.

\bibitem[Andreotti et~al.(2016)Andreotti, B\"aumchen, Boulogne, Daniels,
  Dufresne, Perrin, Salez, Snoeijer, and Style]{Andreottietal16}
B.~Andreotti, O.~B\"aumchen, F.~Boulogne, K.~E. Daniels, E.~R. Dufresne,
  H.~Perrin, T.~Salez, J.~H. Snoeijer, and R.~W. Style.
\newblock Solid capillarity: when and how does surface tension deform soft
  solids?
\newblock \emph{Soft Matter}, 12:\penalty0 2993--2996, 2016.

\bibitem[Bico et~al.(2018)Bico, Reyssat, and Roman]{Bicoetal18}
J.~Bico, E.~Reyssat, and B.~Roman.
\newblock Elastocapillarity: {W}hen surface tension deforms elastic solids.
\newblock \emph{Annual Review of Fluid Mechanics}, 50:\penalty0 629--659, 2018.

\bibitem[Gusev(1997)]{Gusev97}
A.~A. Gusev.
\newblock Representative volume element size for elastic composites: {A}
  numerical study.
\newblock \emph{J. Mech. Phys. Solids}, 45:\penalty0 1449--1459, 1997.

\bibitem[Lopez-Pamies et~al.(2013)Lopez-Pamies, Goudarzi, and Danas]{LPGD13}
O.~Lopez-Pamies, T.~Goudarzi, and K.~Danas.
\newblock The nonlinear elastic response of suspensions of rigid inclusions in
  rubber: {II} --- {A} simple explicit approximation for finite-concentration
  suspensions.
\newblock \emph{J. Mech. Phys. Solids}, 61:\penalty0 19--37, 2013.

\bibitem[Lubachevsky and Stillinger(1990)]{LS90}
B.~D. Lubachevsky and F.~H. Stillinger.
\newblock Geometric properties of random disk packings.
\newblock \emph{J. Stat. Phys.}, 60:\penalty0 561--583, 1990.

\bibitem[Ghosh et~al.(2023)Ghosh, Lef\`evre, and Lopez-Pamies]{GLLP23}
K.~Ghosh, V.~Lef\`evre, and O.~Lopez-Pamies.
\newblock The effective shear modulus of a random isotropic suspension of
  monodisperse liquid $n$-spheres: {F}rom the dilute limit to the percolation
  threshold.
\newblock \emph{Soft Matter}, 19:\penalty0 208--224, 2023.

\bibitem[Lef\`evre et~al.(2022)Lef\`evre, Francfort, and Lopez-Pamies]{LFLP22}
V.~Lef\`evre, G.~{A}. Francfort, and O.~Lopez-Pamies.
\newblock The curious case of 2d isotropic incompressible neo-hookean
  composites.
\newblock \emph{Journal of Elasticity}, 149:\penalty0 1--8, 2022.

\bibitem[Lef\`evre and Lopez-Pamies(2022)]{LLP22}
V.~Lef\`evre and O.~Lopez-Pamies.
\newblock The effective shear modulus of a random isotropic suspension of
  monodisperse rigid $n$-spheres: {F}rom the dilute limit to the percolation
  threshold.
\newblock \emph{Extreme Mechanics Letters}, 55:\penalty0 101818, 2022.

\bibitem[Papanicolaou and Varadhan(1981)]{Papanicolaou81}
G.~{C}. Papanicolaou and S.~{R}.~{S}. Varadhan.
\newblock Boundary value problems with rapidly oscillating random coefficients.
\newblock \emph{Colloquia Mathematica Societatis J\'anos Bolyai}, 27:\penalty0
  835--873, 1981.

\bibitem[Eshelby(1967)]{Eshelby57}
J.~D. Eshelby.
\newblock The determination of the elastic field of an ellipsoidal inclusion
  and related problems.
\newblock \emph{Proc. R. Soc. London A}, 241:\penalty0 376--396, 1967.

\end{thebibliography}

\end{document}